\documentclass[12pt]{iopart}

\usepackage{iopams}  
\begin{document}

\title{On the connection between mutually unbiased bases and orthogonal Latin squares}

\author{T Paterek$^1$, M Paw{\l}owski$^2$, M Grassl$^1$ and {\v C} Brukner$^{3,4}$}

\address{$^1$Centre for Quantum Technologies and Department of Physics,
National University of Singapore, 3 Science Drive 2, 117543 Singapore, Singapore\\
$^2$Institute of Theoretical Physics and Astrophysics,
University of Gda\'nsk, 80-952 Gda\'nsk, Poland\\
$^3$Faculty of Physics, University of Vienna,
Boltzmanngasse 5, A-1090 Vienna, Austria\\
$^4$Institute for Quantum Optics and Quantum Information,
Austrian Academy of Sciences, Boltzmanngasse 3, A-1090 Vienna,
Austria}
\begin{abstract}
We offer a piece of evidence that the problems of finding the number
of mutually unbiased bases (MUB) and mutually orthogonal Latin squares
(MOLS) might not be equivalent.  We study a particular procedure which
has been shown to relate the two problems and generates complete sets of MUBs in
power-of-prime dimensions and three MUBs in dimension six.  For these
cases, every square from an augmented set of MOLS has a corresponding
MUB.  We show that this no longer holds for certain composite
dimensions.
\end{abstract}


\section{Introduction}

Mutually unbiased bases (MUB) encapsulate the concept of
complementarity in quantum formalism.  Quantum observables associated
with these bases are maximally complementary in the sense that given a
system in an eigenstate of one observable, measurement outcomes of the
other observables are completely random.  Although the complementarity
is a distinguishing feature of quantum mechanics, we still do not know
what the total number of mutually maximally complementary observables
for a general $d$-level system is.

It is known that for $d$ being a power of a prime,
there are $d+1$ MUBs \cite{IVANOVIC,WF}
and this number sets the upper bound for arbitrary $d$.
For all other dimensions, it is a puzzle whether this bound is saturated.
Solving this problem gives insight not only into physics,
but also to mathematics, as the problem is linked with other
unsolved mathematical problems \cite{BOYKIN2005, BENGTSSON2007}.
It was also noticed that it is similar in spirit
to some combinatorial problems \cite{ZAUNER, BENGTSSON2004, WB, PDB2008} and problems in finite geometry \cite{SPR,WOOTTERS2006}.
Here, we shall briefly review and study one such connection \cite{PDB2008}.

The problem which is assumed to be connected to finding the number of
MUBs is that of finding the number of mutually orthogonal Latin
squares (MOLS).  The latter has a long history originating in the
works by Euler \cite{EULER} and more is known about it than about the
number of MUBs. 
For example, it is known that there are no more than three squares in the augmented set of MOLS of order six \cite{TARRY}, 
but the question whether or not there are more than three MUBs in dimension six is still open. 
The connection of Ref. \cite{PDB2008} allows to link
every square in an augmented set of MOLS with a MUB for
power-of-prime dimensions and dimension six.  Here we show that this
connection fails in composite dimensions for which MacNeish's
bound is not tight \cite{MACNEISH}.
We study in detail the case of $d=10$,
being the smallest $d$ with this property:
while there are at least four squares in an augmented set of MOLS, 
one cannot find more than three MUBs using the link of Ref. \cite{PDB2008}.

\section{The connection} 

A Latin square of order $d$ is an array of numbers $\{0,...,d-1\}$
where every row and every column contains each number exactly once.
Two Latin squares, $A=[A_{ij}]$ and $B=[B_{ij}]$, are orthogonal if
all ordered pairs $(A_{ij},B_{ij})$ are distinct.  There are at most
$d-1$ MOLS, and such a set of MOLS is called complete.  Complete sets
of MOLS are known to exist for $d$ being a power of a prime.  It is
also known that there are no two MOLS in dimension six \cite{TARRY}.
The existence of $L$ MOLS is equivalent to the existence of a
combinatorial design called a \emph{net} with $L+2$ rows
\cite{DESIGN}.  The net design has a form of a table in which every
row contains $d^2$ distinct numbers.  They are grouped into $d$ cells of
$d$ numbers each, in such a way that the numbers of any cell in a
given row are distributed among all cells of any other row.  The
additional two rows of the net design correspond to orthogonal but not
Latin squares, with the entries $A_{ij} = j$ and $A_{ij} = i$.  
The set of all $L+2$ squares is referred to as the augmented set of MOLS.
An algorithm to construct the design from a set of MOLS is given, e.g.,
in Ref. \cite{PDB2008}.

The MUBs are constructed using the entries of any cell of the design.
We write the entries in modulo $d$ decomposition such that each of
them is now represented by two integers: $m$ and $n$, having values
from the set $\{0,1,\ldots ,d-1\}$.  These integers are taken as
exponents of Weyl-Schwinger operators, $\hat X_d^m \hat Z_d^n$, defined
as:
\[
\hat Z_d | \kappa \rangle = \eta_d^{\kappa} | \kappa \rangle, \qquad
\hat X_d | \kappa \rangle = | (\kappa + 1) \bmod d \rangle,
\]
with $\eta_d = \exp{(i 2 \pi/d)}$ being a complex $d$th root of unity.
The Weyl-Schwinger operators span an unitary operator basis which is
orthogonal with respect to trace scalar product. If one can find sets
of $d$ of them which commute (each set including identity) the joint
eigenbases of the commuting operators form MUBs
\cite{BANDYOPADHYAY,GRASSL2004}.  It turns out that for prime $d$ and
for dimension six, the $d$ operators having exponents from a single
cell of the design commute and therefore define MUBs \cite{PDB2008}.
For power-of-prime $d = p^r$, the operators having exponents from some
cells do not commute.  In order to obtain a complete set of $d+1$ MUBs
one needs to take advantage of the fact that $d$ can be factored.  In
this case, every integer $m$ and $n$ from the net design can be
represented by $r$ digits having their values from the set
$\{0,1,\ldots ,p-1 \}$, i.e. there is a mapping\footnote{Note that
 there are many such mappings. Assuming that $m$ and $n$ can be
 decomposed independently gives $(d!)^2$ maps.} $m \mapsto (m_1,m_2,
\ldots, m_r)$ and $n \mapsto(n_1,n_2,\ldots,n_r)$.  We take these
integers as exponents of tensor product operators $\hat X_p^{m_1} \hat
Z_p^{n_1} \otimes \hat X_p^{m_2} \hat Z_p^{n_2} \otimes \ldots \otimes
\hat X_p^{m_r} \hat Z_p^{n_r}$. For suitable decompositions, related
to finite fields, we find that the operators having exponents from a
single cell of the design again commute and hence define MUBs.

\section{MacNeish's bound}

MacNeish gave a lower bound on the number of MOLS \cite{MACNEISH}.  If
two squares of order $a$ are orthogonal, $A \perp B$, and two squares
of order $b$ are orthogonal, $C \perp D$, then the squares obtained by
a direct product, of order $a b$, are also orthogonal, $A \times C
\perp B \times D$.  This implies that the number of MOLS,
$\mathcal{L}$, of order $d = p_1^{r_1}\ldots p_n^{r_n}$, $p_i$
being prime factors of $d$, is at least $\mathcal{L} \ge
\min_{i}(p_i^{r_i}-1)$, where $p_i^{r_i}-1$ is the number of MOLS of
order $p_i^{r_i}$.

A parallel result holds for MUBs \cite{GRASSL2004,KR}, which we
call the quantum MacNeish bound.  If $| a \rangle$ and $| b \rangle$
are the states of two different MUBs in dimension $d_1$, and $| c
\rangle$ and $| d \rangle$ are the states of two MUBs in dimension $d_2$,
then the tensor product bases $| a \rangle \otimes | c \rangle$ and $|
b \rangle \otimes |d \rangle$ form MUBs in dimension $d_1 d_2$.  Thus,
for $d = p_1^{r_1}\ldots p_n^{r_n}$ there are at least
$\min_{i}(p_i^{r_i}+1)$ MUBs.

Our motivation to study dimension ten comes from the fact that it is
the simplest case in which MacNeish's bound is not tight.
There are at least two MOLS of order ten, which is larger than MacNeish's bound of one. 
If the connection established in Ref. \cite{PDB2008} holds generally, we
shall correspondingly expect the quantum MacNeish bound
not to be tight for $d=10$.  It is already known that the quantum bound is not
tight in general, but the smalltest case for which it was proven is $d
= 26^2 = 676$ \cite{WB}.

\section{Ten dimensions}

The two MOLS of order ten read \cite{SQUARES10}:
\[
\begin{array}{cccccccccc}
0 & 1 & 2 & 3 & 4 & 5 & 6 & 7 & 8 & 9 \\
1 & 2 & 6 & 5 & 8 & 0 & 9 & 3 & 4 & 7 \\
2 & 9 & 4 & 0 & 5 & 7 & 3 & 8 & 6 & 1 \\
3 & 4 & 9 & 7 & 6 & 8 & 5 & 1 & 0 & 2 \\
4 & 3 & 7 & 8 & 1 & 6 & 0 & 2 & 9 & 5 \\
5 & 8 & 3 & 6 & 2 & 9 & 7 & 0 & 1 & 4 \\
6 & 5 & 1 & 9 & 7 & 3 & 8 & 4 & 2 & 0 \\
7 & 0 & 5 & 2 & 9 & 1 & 4 & 6 & 3 & 8 \\
8 & 7 & 0 & 4 & 3 & 2 & 1 & 9 & 5 & 6 \\
9 & 6 & 8 & 1 & 0 & 4 & 2 &5 & 7 & 3
\end{array}
\qquad \quad
\begin{array}{cccccccccc}
0 & 2 & 4 & 9 & 1 & 8 & 7 & 5 & 3 & 6 \\
1 & 7 & 3 & 4 & 5 & 9 & 2 & 6 & 0 & 8 \\
2 & 3 & 8 & 7 & 6 & 4 & 1 & 9 & 5 & 0 \\
3 & 9 & 5 & 2 & 4 & 7 & 0 & 8 & 6 & 1 \\
4 & 5 & 6 & 1 & 9 & 2 & 8 & 0 & 7 & 3 \\
5 & 6 & 2 & 0 & 8 & 1 & 9 & 3 & 4 & 7 \\
6 & 1 & 7 & 8 & 3 & 0 & 4 & 2 & 9 & 5 \\
7 & 4 & 9 & 3 & 0 & 5 & 6 & 1 & 8 & 2 \\
8 & 0 & 1 & 5 & 7 & 6 & 3 & 4 & 2 & 9 \\
9 & 8 & 0 & 6 & 2 & 3 & 5 & 7 & 1 & 4
\end{array}
\]
Using the algorithm of Ref. \cite{PDB2008}, the representative four
cells of the net design read:
\begin{equation}
\begin{tabular}{cccccccccc}
\hline
$00$ & $01$ & $02$ & $03$ & $04$ & $05$ & $06$ & $07$ & $08$ & $09$ 
\\ \hline
$00$ & $10$ & $20$ & $30$ & $40$ & $50$ & $60$ & $70$ & $80$ & $90$
\\ \hline
$00$ & $11$ & $22$ & $33$ & $44$ & $55$ & $66$ & $77$ & $88$ & $99$
\\ \hline
$00$ & $12$ & $24$ & $39$ & $41$ & $58$ & $67$ & $75$ & $83$ & $96$
\\ \hline
\end{tabular}
\label{DESIGN10}
\end{equation}
where we present pairs of numbers $m \, n$.  Writing the pairs as
exponents of operators $\hat X_{10}^m \hat Z_{10}^n$, the first row
gives ten commuting operators $\hat Z_{10}^n$ and therefore defines
the eigenbasis of $\hat Z_{10}$.  Similarly, the second row gives the
eigenbasis of $\hat X_{10}$, and the third row provides the eigenbasis
of $\hat X_{10} \hat Z_{10}$.  However, the operators corresponding to
the fourth row do not commute, e.g. $[\hat X_{10}^2 \hat Z_{10}^4, \hat
 X_{10}^3 \hat Z_{10}^9] \ne 0$, and in this way we do not improve
upon the quantum MacNeish bound,
which is three for $d=10$.

Similar to the case of $d$ being a power of a prime, one may ask if
there is a decomposition of $m$ and $n$ into pairs $m_1 \,m_2$ and
$n_1 \,n_2$ (with $m_1, n_1 \in \{0,1\}$ and $m_2,n_2 \in
\{0,1,2,3,4\}$), respectively, such that the operators $\hat X_2^{m_1}
\hat Z_2^{n_1} \otimes \hat X_5^{m_2} \hat Z_5^{n_2}$, having their
exponents from the corresponding entries of the rows of
(\ref{DESIGN10}), commute.  However, contrary to the case of
power-of-prime $d$, for $d = d_1 d_2$ with coprime factors, the
problem of finding commuting operators $\hat X_{d_1}^{m_1} \hat
Z_{d_1}^{n_1} \otimes \hat X_{d_2}^{m_2} \hat Z_{d_2}^{n_2}$ is
equivalent to the problem of finding commuting operators $\hat X_d^{m}
\hat Z_d^{n}$.  This is a consequence of the lemma in the appendix,
which states that the tensor product operators and the operators in
$d$ dimensions are related by a permutation.

We checked by grouping the commuting operators $\hat X_{10}^{m} \hat
Z_{10}^{n}$, that their eigenbases lead to at most three MUBs.
Similarly, we verified for $d=35$
that there are at most six MUBs formed by the eigenbases of $\hat X_{35}^{m} \hat Z_{35}^{n}$ \cite{WOJTAS}.
These are independent proofs of special cases of the result by Aschbacher,
Childs and Wocjan \cite{ACW}.  In particular, they showed that the
eigenbases of Weyl-Schwinger operators do not lead to more MUBs than
given by the quantum MacNeish bound.  Therefore, for all $d$ in
which MacNeish's bound on the number of MOLS is not tight, the
connection of Ref. \cite{PDB2008} fails to relate all the squares with MUBs.

\section{Conclusions}

We presented some evidence that the physical problem of the number of
MUBs might not be equivalent to the mathematical problem of the number
of MOLS.  The proof of this statement is still an open question.

\section{Acknowledgments}
This work is supported by the National Research Foundation and Ministry of Education in Singapore,
and by the European Commission, Project QAP (No. 015848).
{\v C}B acknowledges support from the Austrian Science Foundation FWF
within Project No. P19570-N16, SFB and CoQuS No. W1210-N16.
The collaboration is a part of an \"OAD/MNiSW program.

\appendix
\section*{Appendix}
\setcounter{section}{1}

\def\bra#1{\langle#1|}
\def\ket#1{|#1\rangle}
\def\C{{\mathbb C}}

Lemma: \textit{If $d=d_1d_2$ with $\gcd(d_1,d_2)=1$, there exists a permutation
matrix $\hat T$ such that}
\[
\hat T \hat X_d \hat T^{-1}= \hat X_{d_1}\otimes \hat X_{d_2}
\qquad and \qquad
\hat T \hat Z_d^{d_1+d_2} \hat T^{-1}= \hat Z_{d_1}\otimes \hat Z_{d_2}.
\]
\underline{Proof}: Define the permutation\footnote{It follows from
 the Chinese remainder theorem that $\hat T$ is a permutation, and
 this is why $d$ has to have coprime factors.}  matrix $\hat T$ by:
\[
\hat T\ket{j} = \ket{(j \bmod d_1) d_2 + j \bmod d_2} \equiv \ket{j \bmod d_1}\ket{j \bmod d_2} \equiv \ket{j_1}\ket{j_2}.
\]
Hence we have
\begin{eqnarray*}
\hat T \hat X_d \hat T^{-1}
&=&\sum_{j=0}^{d-1} \hat T \ket{(j+1)\bmod d}\bra{j} \hat T^{-1}\\
&=&\sum_{j_1=0}^{d_1-1} \ket{(j_1+1)\bmod d_1}\bra{j_1}\otimes
\sum_{j_2=0}^{d_2-1}\ket{(j_2+1)\bmod d_2}\bra{j_2}\\
&=& \hat X_{d_1}\otimes \hat X_{d_2},
\end{eqnarray*}
and
\begin{eqnarray*}
\hat T \hat Z_d^{d_1+d_2} \hat T^{-1}
&=&\sum_{j=0}^{d-1} \eta_d^{(d_1+d_2)j} \hat T \ket{j}\bra{j} \hat T^{-1}\\
&=&\sum_{j=0}^{d-1}\eta_{d_1}^{j} \eta_{d_2}^{j} \ket{j_1} \bra{j_1} \otimes \ket{j_2} \bra{j_2}\\
&=&\sum_{j_1=0}^{d_1-1}\eta_{d_1}^{j_1} \ket{j_1}\bra{j_1}\otimes
\sum_{j_2=0}^{d_2-1}\eta_{d_2}^{j_2}\ket{j_2}\bra{j_2}\\
&=& Z_{d_1}\otimes Z_{d_2},
\end{eqnarray*}
where we used $\eta_{d_i}^j = \eta_{d_i}^{j \bmod d_i} = \eta_{d_i}^{j_i}$ for $i=1,2$.

\end{document}